\documentclass[showpacs,twocolumn,preprintnumbers,nofootinbib,amsmath,amssymb,aps,prd,floatfix]{revtex4-1}
\pdfoutput=1
\usepackage{graphicx}
\graphicspath{{plots/}}
\usepackage{bm}
\usepackage{url}
\usepackage[hyperindex]{hyperref}
\usepackage{todonotes}
\usepackage[normalem]{ulem}
\usepackage[utf8]{inputenc}

\newcommand{\nua}[1]{\ensuremath{\rlap{\kern-2.5pt\ensuremath{\overset{\scriptscriptstyle(-)}{\phantom{\nu}}}}{\ensuremath{{\nu}_{#1}}}}}

\begin{document}

\title{Neutrino Mass Ordering at DUNE: an Extra $\nu$-Bonus}

\author{Christoph A.\ Ternes$^1$}\email{chternes@ific.uv.es}
\author{Stefano Gariazzo$^1$}\email{gariazzo@ific.uv.es}
\author{Rasmi Hajjar$^1$}\email{rashajmu@alumni.uv.es}
\author{Olga\ Mena$^1$}\email{omena@ific.uv.es}
\author{Michel\ Sorel$^1$}\email{sorel@ific.uv.es}
\author{Mariam Tórtola$^1$}\email{mariam@ific.uv.es}

\affiliation{$^1$~Instituto de F\'isica Corpuscular (CSIC-Universitat de Val\`encia), Paterna (Valencia), Spain}

\begin{abstract}
We study the possibility of extracting the neutrino mass ordering at
the future Deep Underground Neutrino Experiment using atmospheric
neutrinos, which will be available  before the muon neutrino  beam
starts being operational. The large statistics of the atmospheric muon neutrino
and antineutrino samples at the far detector, together with the baselines of thousands
of kilometers that these atmospheric (anti)neutrinos travel, provide ideal ingredients to extract the neutrino mass ordering via matter effects in the neutrino propagation through
the Earth. Crucially, muon capture by Argon provides excellent
charge-tagging, allowing to disentangle the neutrino and
antineutrino signature. This is an important extra benefit of having a Liquid Argon Time Projection
Chamber as far detector, that could render a $\sim 3.5\sigma$ extraction
of the mass ordering after approximately seven years of exposure. 
\end{abstract}


\maketitle

\newpage
\section{Introduction}
\label{sec:intro}
Neutrino oscillation experiments imply the first departure from
the Standard Model (SM) of Particle Physics, as they have found
overwhelming evidence for the existence of neutrino masses. 
Despite the accuracy they provide on the neutrino oscillation
parameters --- which is of the order of the percent
level~\cite{deSalas:2017kay} --- the sign of the atmospheric mass
splitting, $\Delta m_{31}^2$, and the value of the CP
violating phase $\delta$ remain both unknown.
The sign of $\Delta m_{31}^2$ originates two possible scenarios,
\emph{normal} (NO) or \emph{inverted} ordering (IO)~\cite{Gariazzo:2018pei,deSalas:2018bym}. The sensitivity to the neutrino mass spectrum
at oscillation experiments is mostly coming from the presence of
matter effects~\cite{Wolfenstein:1977ue,Mikheev:1986gs,Parke:1986jy,Krastev:1989ix,Akhmedov:1992mm,Petcov:1998su,Chizhov:1998ug,Akhmedov:1998ui,Chizhov:1999az,Chizhov:1999he} in the neutrino and antineutrino propagation.  
In the normal (inverted) mass ordering scenario, the neutrino
flavor transition probabilities will get enhanced (suppressed),
while in the case of antineutrino propagation
the opposite happens and the antineutrino flavor transition probabilities will get suppressed (enhanced) in the normal (inverted) mass ordering
scenario.
At long-baseline accelerator experiments,
matter effects, and consequently the sensitivity to the mass
ordering, increase with the baseline,
while these effects will be negligible at short-baseline and
medium-baseline experiments.
Despite this, when extracting both the mass ordering and the CP violating phase from results of long-baseline facilities, knowledge on the mixing angle $\theta_{13}$ in vacuum is 
required. Short- and medium-baseline experiments at reactors have been fundamental to establish strong constraints on such angle.
Despite the fact that the neutrino mass ordering remains unknown,
current oscillation data are mildly favouring the normal ordering
scenario. The authors of Refs.~\cite{deSalas:2018bym,deSalas:2017kay} have reported a
global preference for normal ordering at the level of $2.7 \sigma$
from all long-baseline accelerator and short-baseline reactor data
(i.e. \texttt{T2K}, \texttt{NO$\nu$A}, \texttt{K2K}, \texttt{MINOS},
\texttt{Daya Bay}, \texttt{RENO} and
\texttt{Double Chooz}).

The future long baseline facility \texttt{DUNE} (Deep Underground Neutrino
Experiment)~\cite{Abi:2018dnh,Abi:2018alz,Abi:2018rgm} aims to extract the sign of the
atmospheric mass splitting and the CP violating phase
$\delta$ through the golden channels $\nu_\mu \to \nu_e$ and
$\bar{\nu}_\mu \to \bar{\nu}_e$, the same channels exploited by the current
\texttt{T2K}~\cite{Abe:2018wpn,t2k} and \texttt{NO$\nu$A}~\cite{NOvA:2018gge,nova} experiments.
However, both quantities can also
be extracted using atmospheric neutrino beams~\footnote{See Ref.~\cite{Razzaque:2014vba} and the recent work of \cite{Kelly:2019itm} for a CP violation measurement using sub-GeV atmospheric neutrinos.}. Indeed, the idea of using atmospheric neutrino fluxes to distinguish the type of mass ordering is well-known in the literature since a long time~\cite{Banuls:2001zn,Bernabeu:2003yp}. These pioneer studies focused mostly on muon calorimeter detectors, such
as \texttt{MONOLITH}~\cite{TabarellideFatis:2002ni},
\texttt{MINOS}~\cite{Sousa:2015bxa} or
\texttt{INO}~\footnote{\url{http://www.ino.tifr.res.in/ino/}}  in
which the muon charge can be determined, see also Refs.~\cite{PalomaresRuiz:2004tk,Indumathi:2004kd,Gandhi:2005wa,Petcov:2005rv,Huber:2005ep,Samanta:2006sj,Gandhi:2007td,Gandhi:2008zs,Samanta:2009qw,Barger:2012fx,Ghosh:2012px,Blennow:2012gj,Ghosh:2013mga,Devi:2014yaa}. 
Furthermore,
in the absence of  a charged current event by event final muon
charge discrimination, the addition of the atmospheric neutrino
oscillation data to the analysis performed in Ref.~\cite{deSalas:2017kay} improves the preference for normal ordering
to the level of 3.4$\sigma$, mostly due to the
\texttt{Super-Kamiokande} $\nu_\mu \to \nu_e$ data
sample~\cite{Abe:2017aap}, where the separation among $\nu_e$ and
$\bar{\nu}_e$ events is done statistically.

Neutrino observatories can also extract the sign of the atmospheric
mass difference with lower energy detection thresholds for atmospheric
neutrino extensions by looking at the less sensitive but higher
statistics muon disappearance channels such as $\nu_\mu \to \nu_\mu$ and
$\bar{\nu}_\mu \to \bar{\nu}_\mu$~\cite{Mena:2008rh}. The IceCube
collaboration has recently reported a preference for NO
with a p-value of $p_{\rm {IO}}=15.3\%$  for the IO
hypothesis~\cite{Aartsen:2019eht} using data collected by the DeepCore extension. This will also be the main target for \texttt{ORCA}~\cite{Adrian-Martinez:2016fdl,Capozzi:2017syc} and
\texttt{PINGU}~\cite{Ge:2013zua,Ge:2013ffa,Aartsen:2014oha,Capozzi:2015bxa}, see e.g.~\cite{Akhmedov:2012ah,Agarwalla:2012uj,Ribordy:2013set,Franco:2013in,Winter:2013ema,Blennow:2013oma,Blennow:2013vta,Aartsen:2013aaa}.

In this manuscript, we exploit the atmospheric neutrino signatures at
the DUNE detector, a  Liquid Argon Time Projection
Chamber (LArTPC). Despite this detection technology, in the absence of a
magnetic field, does not allow for a charge identification of the
final lepton state, one can make use of
a particular event topology available in Argon detectors: muon
capture. This bonus process will provide a clean measurement of  the
muon charge, that will considerably improve the capabilities of DUNE
to perform mass ordering measurements with atmospheric
neutrinos. Notice that the advantage is
twofold, as \textit{(i)} measurements of the mass ordering could be available
before the beam starts, and \textit{(ii)} the combination with the
beam information will notably enhance the expected sensitivity reach. 
We shall show that muon capture events could greatly enhance the sensitivity
to the mass ordering from atmospheric neutrinos only. For an earlier,
and preliminary, appraisal of the neutrino mass ordering sensitivity
in DUNE using atmospheric neutrinos, including a statistical
discrimination between neutrinos and antineutrinos, see
Ref.~\cite{Acciarri:2015uup}. The latter work was largely based on
previous studies in the framework of the LBNE project,
see~\cite{Adams:2013qkq}. 

The structure of the paper is as follows: In Sec.~\ref{sec:matter}, we
describe the oscillations of atmospheric neutrinos and the matter
effects they undergo. Next, in Sec.~\ref{sec:detector}, we discuss the simulation of the neutrino event rates at the DUNE far detector and how the muon capture comes into play. Sec.~\ref{sec:analysis} contains the description of the statistical
method and the main results obtained in this study. Our final remarks are presented in
Sec.~\ref{sec:concl}. 


\section{Matter effects and atmospheric neutrinos}
\label{sec:matter}

In atmospheric neutrino experiments, the
size of matter effects is given by the effective mixing angle
$\theta_{13}$ in matter, which leads to the golden channel transitions 
$\nu_\mu \to \nu_e$, $\nu_e \to \nu_\mu$, $\bar{\nu}_\mu \to
\bar{\nu}_e$ and $\bar{\nu}_e \to
\bar{\nu}_\mu$ and reads, within the simple two-flavor mixing framework, as
\begin{equation}
\sin^2 2 \theta^{\textrm{m}}_{13}
=
\frac{\sin^2 2 \theta_{13}}%
{\sin^2 2 \theta_{13} +
\left(\cos 2 \theta_{13} \mp
\frac{A}{\Delta m_{31}^2}
\right)^2}~,
\label{eq:mixmatter}
\end{equation}
where the minus (plus) sign refers to neutrinos (antineutrinos). The matter potential is given by $A=2 \sqrt{2} G_{F} N_{e} E$ and $N_e$ is the electron number density in the Earth interior. Consequently, 
matter effects will enhance (deplete) the neutrino (antineutrino)
oscillation probabilities $P (\nu_\mu\to \nu_e)$  and $P (\nu_e\to \nu_\mu)$
($P
(\bar{\nu}_\mu\to \bar{\nu}_e)$ and $P
(\bar{\nu}_e\to \bar{\nu}_\mu)$) if the mass ordering is normal. When
the resonance condition 
\begin{equation}
\Delta m^2_{31} \cos 2 \theta_{13}
=
2 \sqrt{2} G_{F} N_{e} E
\label{eq:res}
\end{equation}
is satisfied, matter effects are expected to have their largest contribution. 
In the case of atmospheric neutrinos, which travel distances of several
thousand of kilometers, and for $\Delta m^2_{31} \sim 2.5 \times
10^{-3}$~eV$^2$~\cite{deSalas:2017kay}, the resonance condition will take place at neutrino energies
$\sim 3-8$~GeV, depending on the precise value of $N_{e}$ in the
Earth's interior. 

\begin{figure}
\centering
\begin{tabular}{c c}
\includegraphics[width=0.25\textwidth]{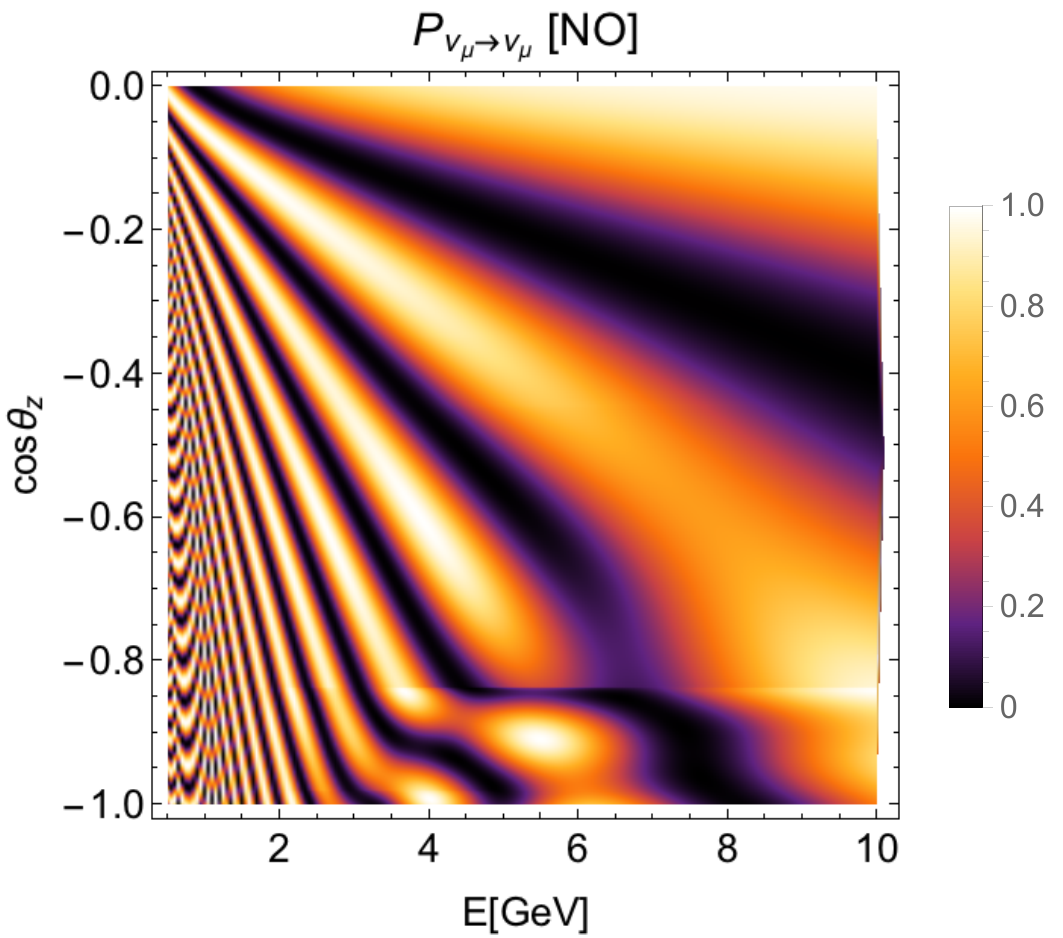} & \includegraphics[width=0.25\textwidth]{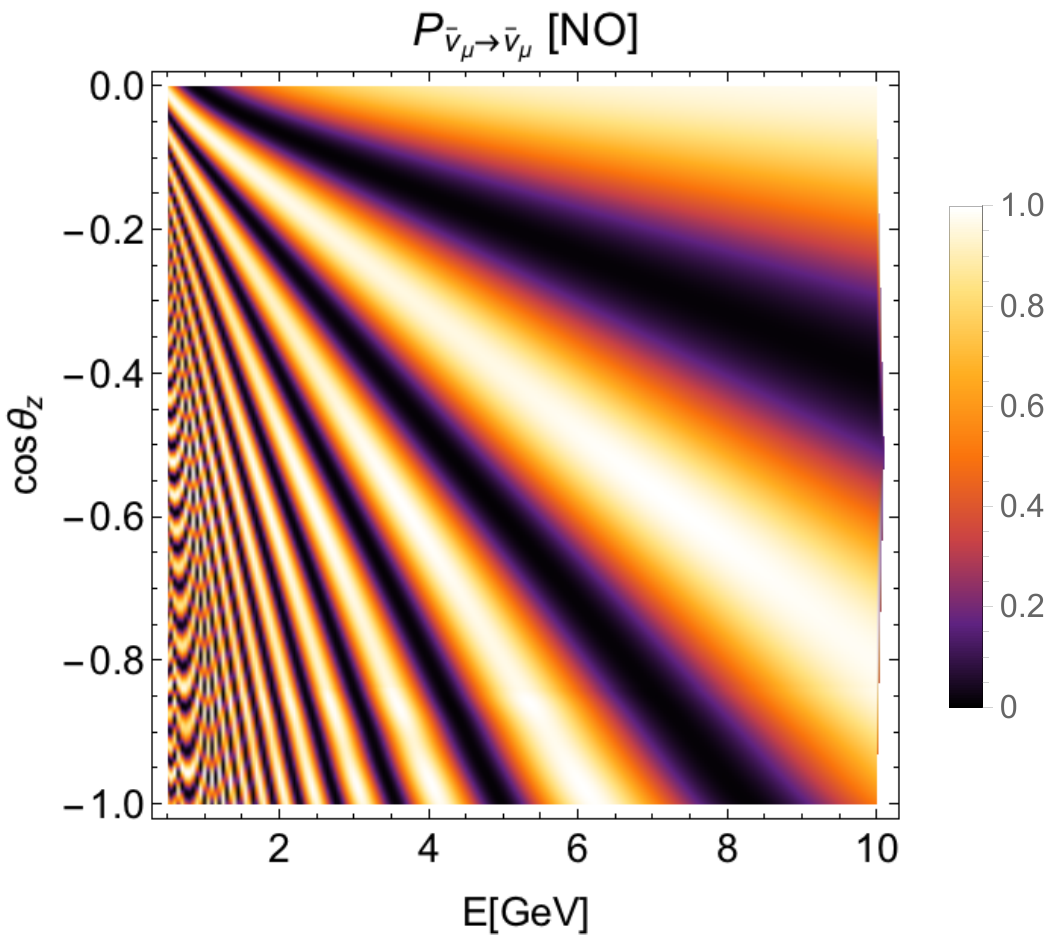}\\
\includegraphics[width=0.25\textwidth]{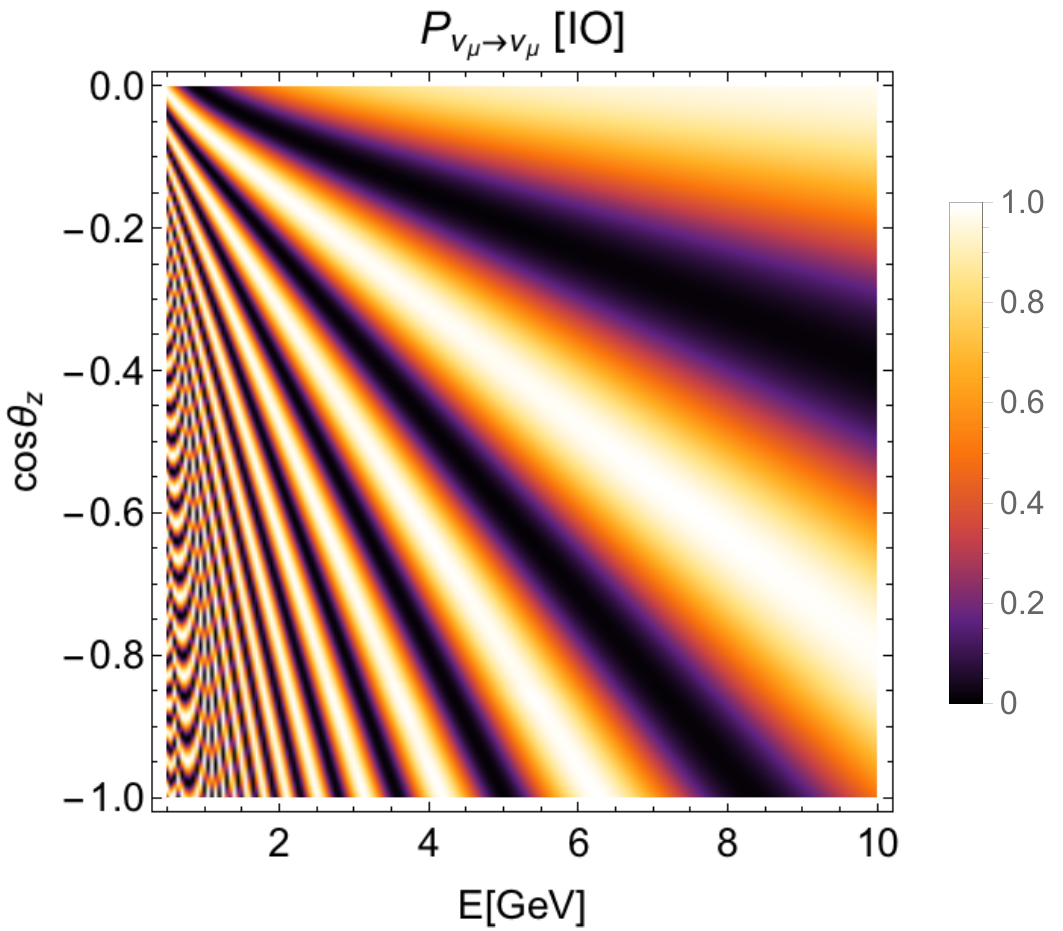}
& \includegraphics[width=0.25\textwidth]{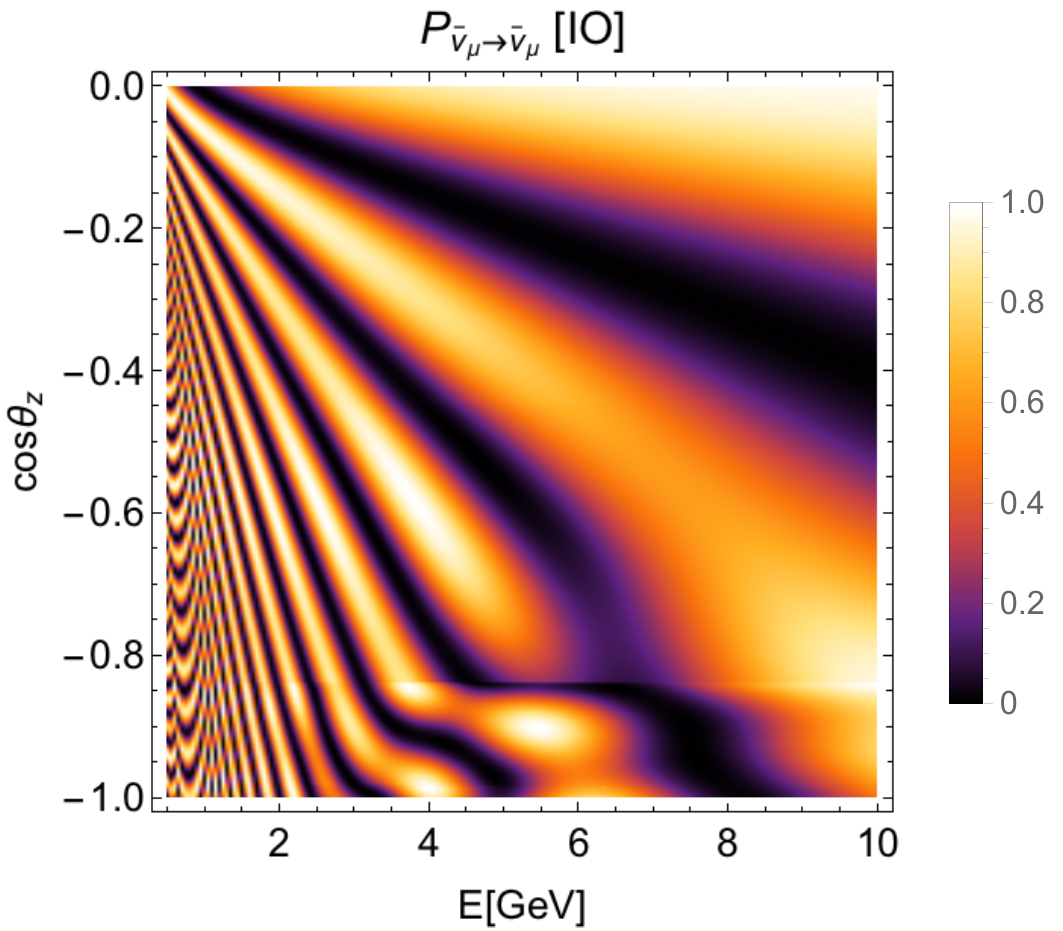}
\end{tabular}
\caption{\label{fig:Pmumu} Left panels:
survival probability $P (\nu_\mu\to \nu_\mu)$ 
as a function of the neutrino energy $E$ and
the cosine of the zenith angle, $\cos\theta_z$,
for normal (inverted) ordering in the top (bottom) line. Right panels:
same as in the left panels, but  for the antineutrino channel.}
\end{figure}

Matter effects are also present in the muon disappearance channels
$P (\nu_\mu\to \nu_\mu)$ and $P (\bar{\nu}_\mu\to \bar{\nu}_\mu)$,
relevant for both long-baseline and atmospheric neutrino beams. In the simplified case of a constant matter density,
the disappearance probability at terrestrial baselines~\footnote{For an
expansion with solar mixing effects included, see Ref.~\cite{Akhmedov:2004ny}.} is given by
\begin{widetext}
\begin{eqnarray}
P\left(\nua{\mu} \to \nua{\mu}\right)
&=&
1-\cos^2 \theta^{\textrm{m}}_{13}
\sin^2 2 \theta_{23}
\times
\sin^2\left[1.27\left(\frac{\Delta m^2_{31} + A +
(\Delta m^2_{31})^{\textrm{m}}}{2}\right)
\frac{L}{E}\right]
\label{eq:mixmatter2}
\\
&&
-\sin^2 \theta^{\textrm{m}}_{13}
\sin^2 2 \theta_{23}
\times
\sin^2\left[1.27\left(\frac{\Delta m^2_{31} + A -
(\Delta m^2_{31})^{\textrm{m}}}{2}\right)
\frac{L}{E}\right]
- \sin^4\theta_{23}\sin^2 2 \theta^{\textrm{m}}_{13}
\sin^2\left[1.27 (\Delta m^2_{31})^{\textrm{m}}\frac{L}{E}\right]
~ \nonumber,
\end{eqnarray}
\end{widetext}
where
$\theta^{\textrm{m}}_{13}$ is that of Eq.~\eqref{eq:mixmatter}
and
\begin{equation}
(\Delta m^2_{31})^{\textrm{m}}
=
\Delta m^2_{31}
\sqrt{\sin^2 2 \theta_{13}
+ \left( \cos 2 \theta_{13} \mp
\frac{A}{\Delta m^2_{31}}
\right)^2}~.
\label{eq:masseff}
\end{equation}
The muon survival probabilities will be suppressed (enhanced)  if the ordering is normal (inverted),
so the effect is opposite to the one present in the $\nu_e \to
\nu_\mu$ oscillation channel. Therefore, when dealing with atmospheric neutrino beams, since there
is an irreducible muon neutrino background from $\nu_e \to
\nu_\mu$ oscillations, the size of the matter effects will be
reduced. The distance $L$ traveled through the Earth by these
atmospheric neutrino beams is fixed by their arrival zenith angle $\theta_z$ (with $\cos
\theta_z=1$ for vertical down-going neutrinos and $\cos
\theta_z=-1$ for vertical up-going neutrinos),
\begin{equation}
L=R_\oplus \left(\sqrt{\left(1+\frac{h}{R_\oplus}\right)^2 -(1-\cos
\theta_z)^2}- \cos
\theta_z \right)~,
\end{equation}
with $R_\oplus$ the Earth's radius and $h\simeq 15$~km the neutrino production distance
from the Earth's surface. The dependence of the survival probabilities
$P (\nu_\mu\to \nu_\mu)$ and $P (\bar{\nu}_\mu\to \bar{\nu}_\mu)$ on
the neutrino energy $E$ and the cosine of the zenith angle, $\cos\theta_z$, is shown in the left and right panels of Fig.~\ref{fig:Pmumu} for normal 
and inverted ordering (top and bottom figures). Notice that, in the case of normal
ordering, the resonance takes place at the aforementioned energies
(3-8 GeV) for almost vertical up-going neutrinos,
$-1<\cos\theta_z<-0.8$, while for the inverted ordering, such a
resonant enhancement in the transition probabilities will take place
in the antineutrino channel instead. Therefore, even if both the angular
and the energy resolution of the detector should be optimal, the key
ingredient to disentangle matter effects (and, ultimately, the neutrino
mass ordering) is to have a detector with muon charge tagging,
generally achieved with a magnetized
detector. However,  as we shall shortly see, LArTPCs allow for such a
possibility without the need of a magnetic field.

\section{Atmospheric neutrino events in DUNE: muon capture in Argon}
\label{sec:detector}
Our statistical analyses will deal with three possible fully contained event samples at atmospheric
neutrino detectors: $\mu^-$-like events that undergo muon capture ($N_{i,j,\mu}^\text{cap}$), 
the rest of the muons and all of the antimuons that undergo muon decay ($N_{i,j,\mu}^\text{rest}$), and $e$-like events 
($N_{i,j,e}$)~\footnote{Electron charge identification is impossible at GeV energies and we shall consider just one event sample
which accounts for both $e^+$ and $e^-$-like events.}.

Let us start with the $\mu^-$-like contained events 
produced by the interactions of atmospheric up-going neutrinos in the
LArTPC DUNE detector. In a LArTPC, both ionization charge~\cite{Cavanna:2014iqa,Acciarri:2017sjy} and scintillation light~\cite{Sorel:2014rka} information can be used to infer the neutrino/antineutrino content in a muon neutrino beam. This is possible by exploiting the signature of $\mu^-$ capture on Argon nuclei, only available for
contained events. In argon, the effective $\mu^-$ lifetime resulting from the competing decay and nuclear capture processes is given by:
\begin{equation}
  \tau = (1/\tau_\text{cap} + Q/\tau_\text{free})^{-1}
\end{equation}
where $\tau_\text{cap}$ is the lifetime of the capture process, $Q=0.988$ is the Huff correction factor~\cite{Suzuki:1987jf} and $\tau_\text{free} = 2197.0$~ns~\cite{Tanabashi:2018oca} is the muon lifetime in vacuum. The resulting $\mu^-$ capture fraction is then given by:
\begin{equation}
  \epsilon^\text{cap} = \tau/\tau_\text{cap} = 1 - \tau/\tau_\text{free}
\end{equation}
  
The most precise determination of the $\mu^-$ lifetime in argon was obtained in \cite{Klinskikh2008}, $\tau = (616.9 \pm 6.7)$~ns, resulting in 
\begin{equation}
  \epsilon^\text{cap} = (71.9\pm 0.3)\%
  \label{eq:muminuscapture}
\end{equation}
This measurement is fully compatible with the earlier measurement of $\tau = (606 \pm 29)$~ns in \cite{Suzuki:1987jf} and the preliminary result from LArIAT of $\tau = (626 \pm 48)$~ns \cite{Foreman:2019bjd}. In our analysis, we use the central value and uncertainty in Eq.~\eqref{eq:muminuscapture}.

For our sensitivity estimates, we also assume a 100\% efficiency for tagging Michel electrons and positrons from $\mu^{\pm}$ decays at rest, as done in \cite{Adams:2013qkq}. Any tagging inefficiency would cause decay events to be mis-interpreted as capture events, and should therefore be avoided for optimal muon neutrino/antineutrino separation. We consider this approximation to be sufficient for the purposes of this feasibility study. Efficiency estimates using detailed DUNE simulations are not yet publicly available. Still, early data from ICARUS \cite{Amoruso:2003sw} and LArIAT \cite{Foreman:2019bjd} have already shown that the Michel electron tagging efficiency can reach values close to unity in LArTPC detectors using either charge or light information. In any case, any Michel electron tagging inefficiency smaller than $(1-\epsilon^\text{cap})\simeq 28\%$ will have a sub-dominant contribution to the mixing  of muon neutrino and muon antineutrino stopping samples in our analysis, compared to the effect of $\mu^-$ tracks that do not capture and decay.

Therefore, it appears
possible to select a statistically significant, highly pure, sample of
$\mu^-$-like atmospheric neutrino interactions, with an identification
efficiency of $\epsilon^\text{cap}$, as given in Eq.~\eqref{eq:muminuscapture}.
The number of muon-like contained
events in the $i$-th neutrino energy ($E_r$) and $j$-th cosine of the zenith angle
($c_{r,\nu}$) bin (both reconstructed quantities) reads as
\begin{widetext}
\begin{eqnarray}
N_{i,j,\mu^-(\mu^+)}&=& \frac{2 \pi N_{\rm T} \, t}{V_{\rm det}} \,
\int_{E_{r,i}}^{E_{r,i+1}}dE_{r,\nu} \int_{c_{r,\nu,
j}}^{c_{r,\nu,j+1}}
dc_{r,\nu} 
\int_{0}^\infty dE_{\nu} \int_{-1}^{1} 
dc_{\nu} V_\mu
\nonumber
\\
&&
\times
\left(\frac{d\phi_{\nu_e(\nu_\mu) (\bar{\nu}_e(\bar{\nu}_\mu))}}{dE_\nu d\Omega}\, \sigma^{\rm CC}_{\nu_\mu(\bar{\nu}_\mu)}
P_{\nu_e(\nu_\mu) \to \nu_\mu (\bar{\nu}_e(\bar{\nu}_\mu) \to \bar{\nu}_\mu) }\right)
\times R^\mu_e(E_{r,\nu}, E_\nu) R^\mu_\theta (\theta_{r,\nu},\theta_\nu)  ~,
\label{eq:events}
\end{eqnarray}
\end{widetext}
where $d\phi_\nu$'s are the atmospheric neutrino differential
fluxes, $\sigma^{\rm CC}$ is the CC neutrino cross sections in Argon, $N_{\rm T}$ is the number of available targets, $V_{\rm det}$ is the total volume of the detector, $V_{\mu}$ is the effective detector volume and $t$ is the exposure
time. Finally, $R^\mu_e(E_{r,\nu}, E_\nu)$ and $R^\mu_\theta
(\theta_{r,\nu},\theta_\nu)$ account for the energy and angular
smearing. 

The $\mu^-$-like
contained events that undergo muon capture are given by
\begin{equation}\label{eq:Ncap}
N_{i,j,\mu}^\text{cap} = \epsilon^{\text{cap}}\,N_{i,j,\mu^-}\,,
\end{equation}
while the remaining muon-like events are given by 
\begin{equation}\label{eq:Nrest}
N_{i,j,\mu}^\text{rest} = (1 - \epsilon^{\text{cap}})\,N_{i,j,\mu^-} + N_{i,j,\mu^+}\,.
\end{equation}

\begin{widetext}

\begin{figure}
  \centering
  \includegraphics[width=0.32\textwidth]{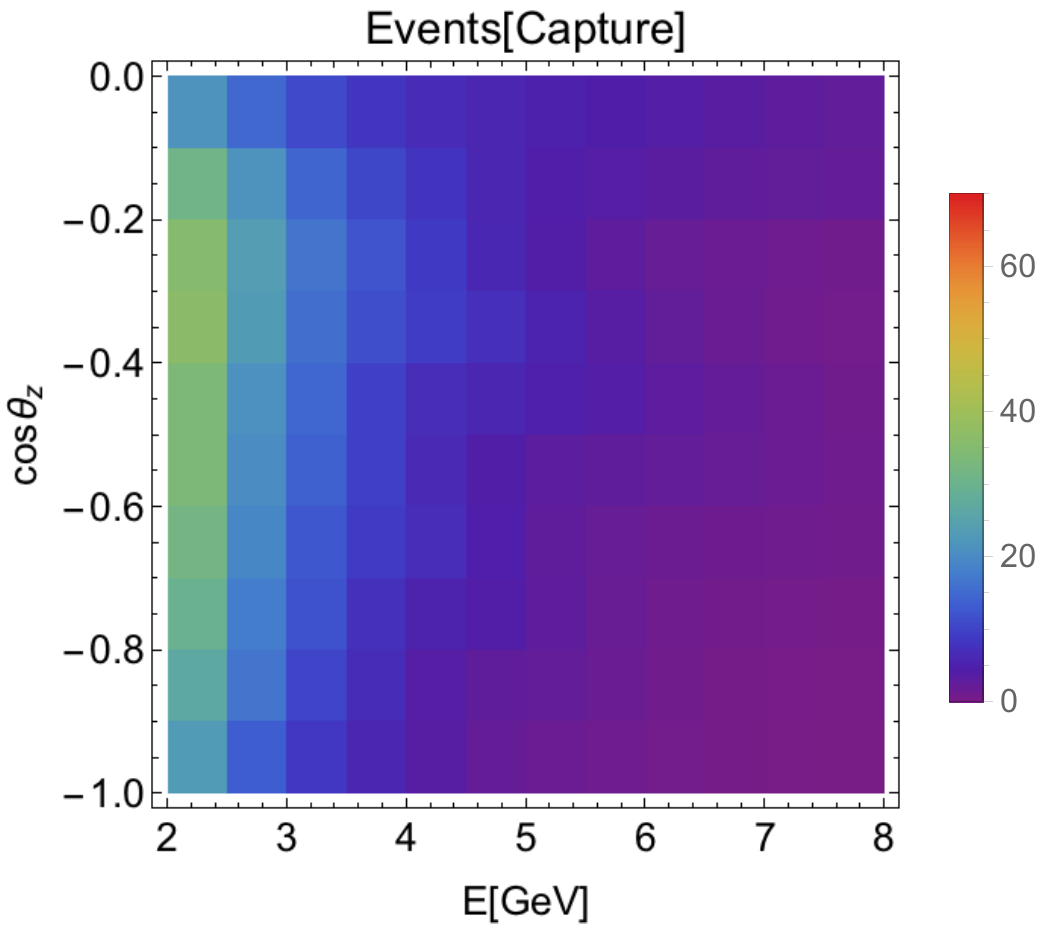}
  \includegraphics[width=0.32\textwidth]{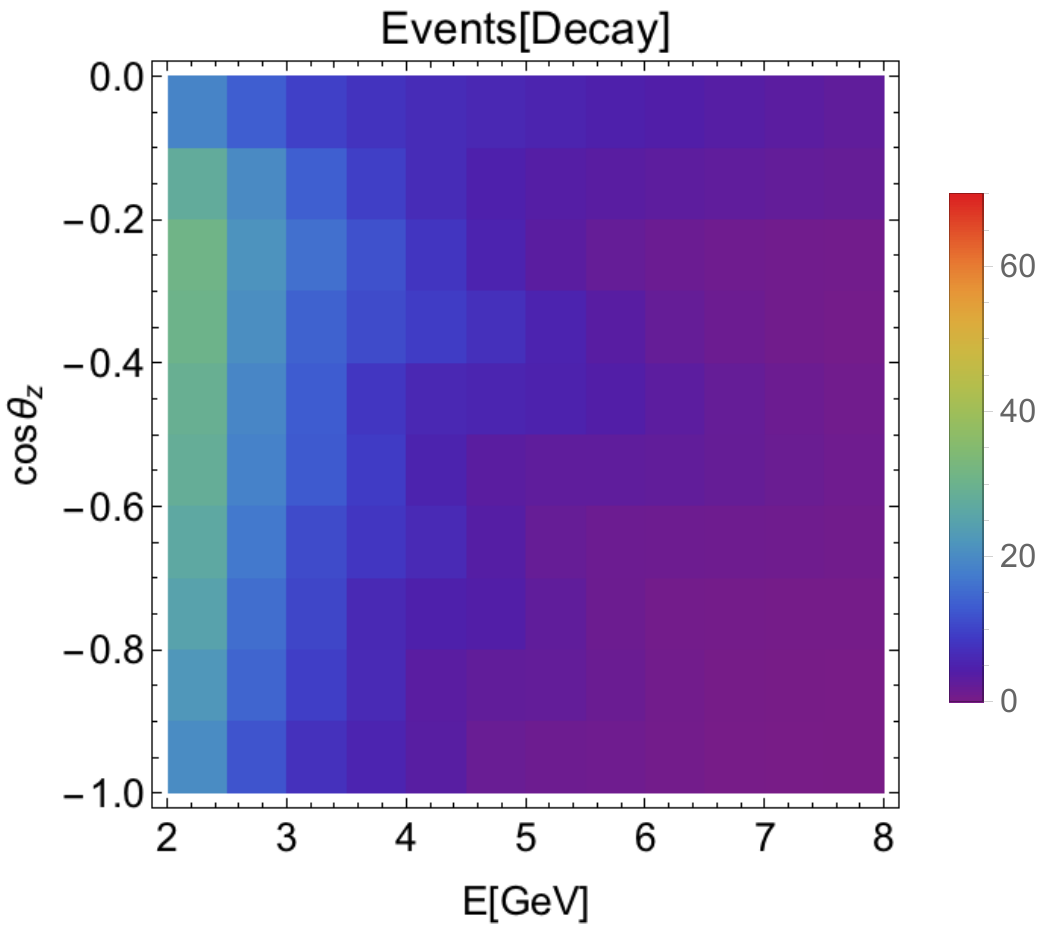}
  \includegraphics[width=0.32\textwidth]{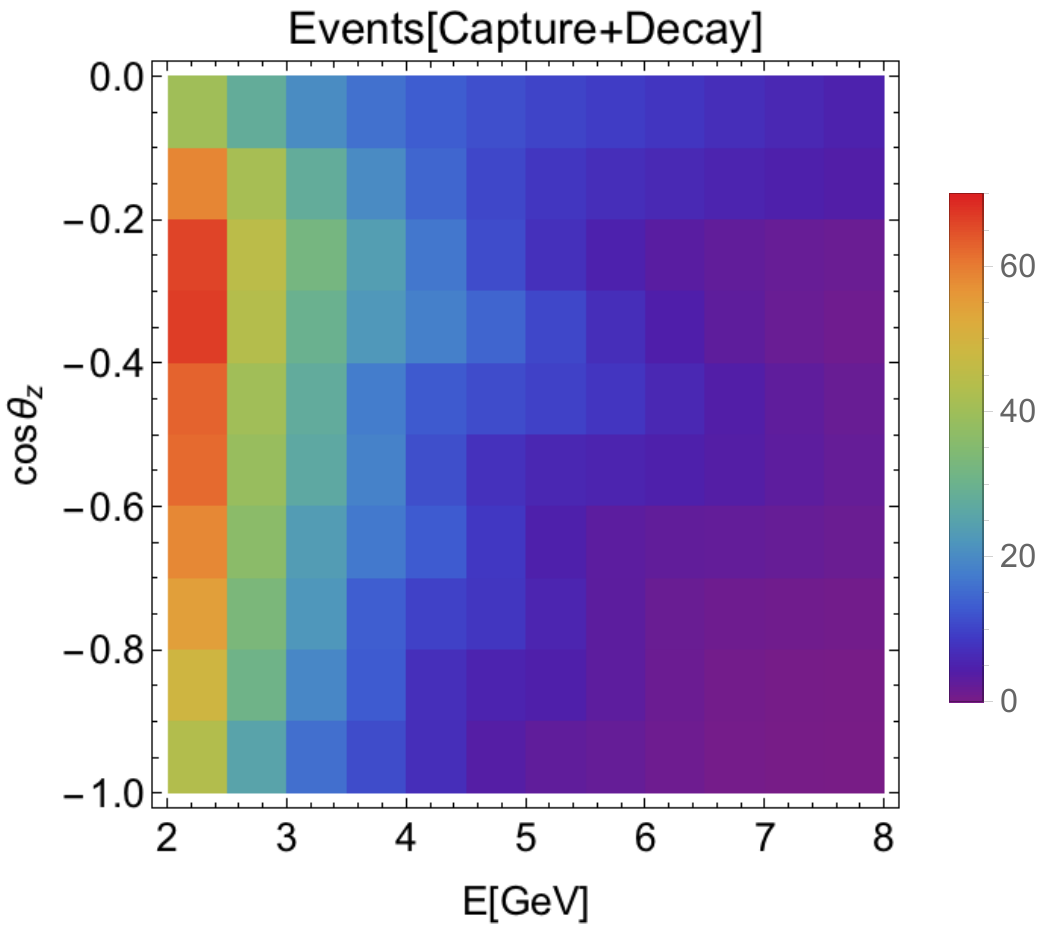}
  \caption{Number of expected muon events after 400~kt$\cdot$yr exposure time, separating capture and decay events (first two panels) and combining all muon type events (right panel), using the parameters in Tab.\ref{tab:params} and $\sin^2\theta_{23} = 0.547$.
  }
  \label{fig:events}
\end{figure}
\end{widetext}

In Fig.~\ref{fig:events} we show an example for the expected number of events, fixing the value the oscillation parameters to the ones in Tab.~\ref{tab:params} and the atmospheric mixing angle to $\sin^2\theta_{23} = 0.547$. In the first two panels, we plot the capture and decay events separately, respectively, while showing the combination of both samples in the right panel. Note that when trying to reconstruct the right panel there can be more degeneracies among parameters in the analysis than when fitting the two sets independently and, therefore, one expects to obtain stronger results. This is indeed the case, as we will see below.

In the case of electrons, the number of $e$-like events in the
$i$-th and $j$-th bin in ($E_r,\, c_{r,\nu}$) reads as
\begin{widetext}
\begin{eqnarray}
N_{i,j,e^-(e^+)}&=& 2 \pi N_{\rm T} \, t \,
\int_{E_{r,i}}^{E_{r,i+1}}dE_{r,\nu} \int_{c_{r,\nu,
j}}^{c_{r,\nu,j+1}}
dc_{r,\nu} 
\int_{0}^\infty dE_{\nu} \int_{-1}^{1} 
dc_{\nu}
\nonumber
\\
&&
\times\left(\frac{d\phi_{\nu_e(\nu_\mu) (\bar{\nu}_e(\bar{\nu}_\mu))}}{dE_\nu d\Omega}\, \sigma^{\rm CC}_{\nu_e(\bar{\nu}_e)}
P_{\nu_e(\nu_\mu) \to \nu_e (\bar{\nu}_e(\bar{\nu}_\mu) \to \bar{\nu}_e) }\right)
\times R^e_e(E_{r,\nu}, E_\nu) R^e_\theta (\theta_{r,\nu},\theta_\nu)~.
\label{eq:eventse}
\end{eqnarray}
\end{widetext}

As previously stated, we just consider one electron-like event sample 
$N_{i,j,e}$ which is computed as the sum of $N_{i,j,e^-}$ and
$N_{i,j,e^+}$.

Regarding the atmospheric electron and muon (anti) neutrino fluxes,
for the differential fluxes
$\frac{d\phi_{\nu_\alpha}}{dE_\nu d\Omega}$ that appear in Eqs.~(\ref{eq:events}) and (\ref{eq:eventse}),
we use the results from Ref.~\cite{Barr:2004br}, albeit
very similar numbers would have been obtained using the fluxes from
Refs.~\cite{Honda:2006qj,Honda:2011nf,Honda:2015fha}. We shall
comment in the following section on the errors on these
atmospheric neutrino fluxes, that have been properly added to other
sources of systematic uncertainties in our numerical studies. 

The cross sections for muon and electron (anti)neutrino interactions on Argon nuclei  in the 0--10~GeV neutrino energy range have been simulated by means of the GENIE Monte Carlo neutrino event generator~\cite{Andreopoulos:2009rq}. GENIE is extensively used by the neutrino physics community and by the DUNE Collaboration in particular. As our cross section model, we use the total charged-current (anti)neutrino cross sections provided by GENIE version 2.12.10 on $^{40}$Ar nuclei (18 protons and 22 neutrons). The model accounts for a comprehensive list of interaction processes, including quasi-elastic scattering, baryon resonance production, coherent pion production in neutrino-nucleus scattering and deep inelastic scattering. Nuclear effects affecting  total cross sections are included. Final state hadronic interactions occurring within the Argon target nucleus are not simulated, but indirectly accounted for via our assumed energy and angular resolution functions. 

To compute the effective volume fraction $V_{\mu}/V_{\rm det}$ in Eq.~\eqref{eq:events} for contained muon events, we have
approximated the DUNE detector to be made of four independent modules
with approximately 13~kton of LAr active mass each, 
each of them assumed to have an elliptical
cylindrical shape of 12~m height and major and minor axis of $a=29$~m
and $b=7.25$~m, respectively. For the calculation of the effective
volume we have taken into account the muon range in Argon
$R_\mu(E_\mu)$, which depends on the lepton energy.  
Conservatively, we have also computed the number of $\mu^+$-like
events restricting ourselves to the contained topology. This
assumption eases the comparison with respect to the case in which no
flavor tagging is available and ensures
good energy reconstruction for the full muon like event sample.

\begin{figure}
  \centering
  \includegraphics[width=0.5\textwidth]{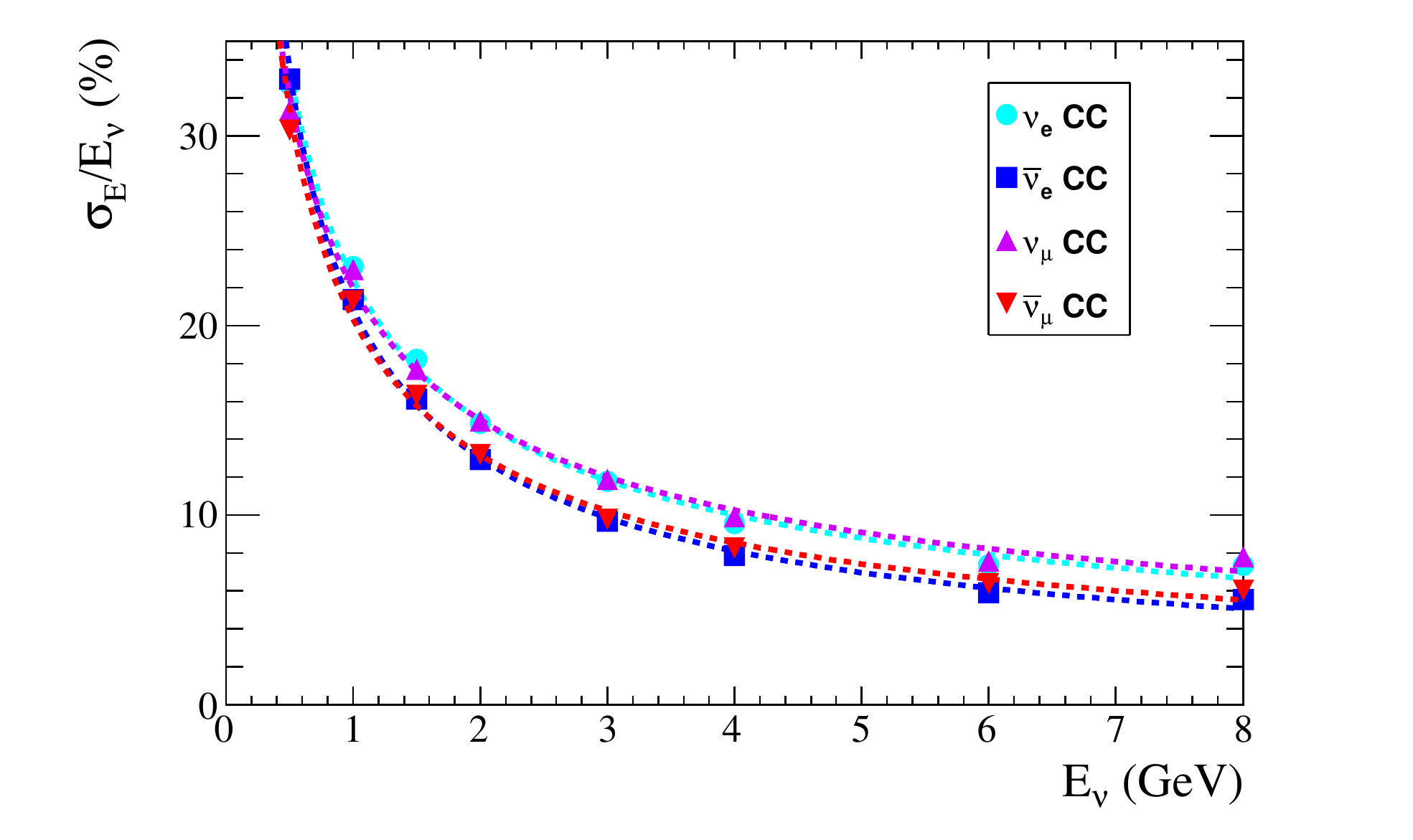}
  \includegraphics[width=0.5\textwidth]{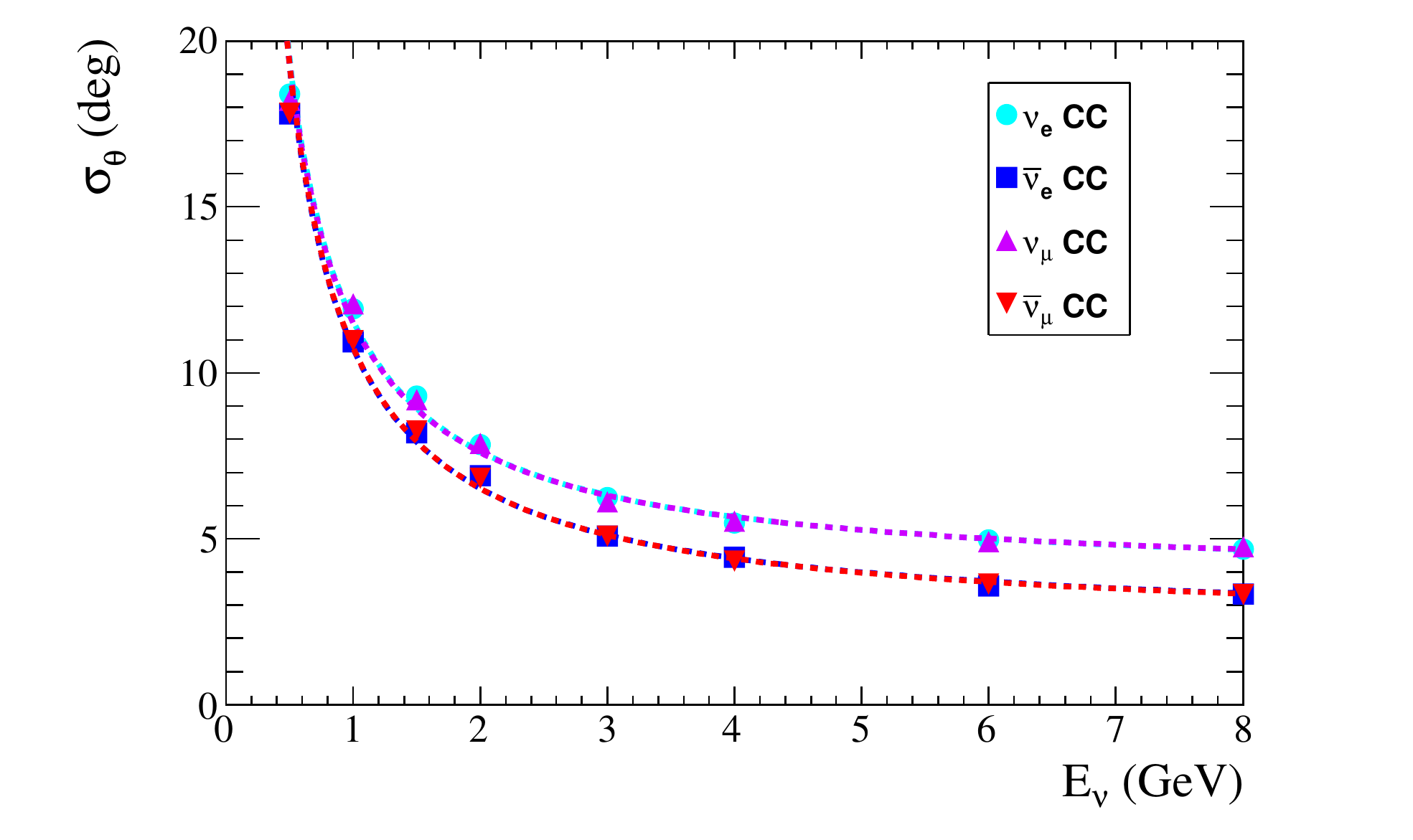}
  \caption{Relative neutrino energy resolution $\sigma_E/E_{\nu}$ (top) and absolute neutrino angular resolution $\sigma_\theta$ (bottom) as a function of neutrino energy $E_{\nu}$ assumed in this study, for $\nu_e$, $\bar{\nu}_e$, $\nu_\mu$ and $\bar{\nu}_\mu$ charged-current interactions on argon nuclei.}
  \label{fig:resolutions}
\end{figure}

As for the energy and angular smearing inherent to reconstruction processes and final state hadronic interactions within Argon nuclei, $R^\mu_e(E_{r,\nu}, E_\nu)$ and $R^\mu_\theta(\theta_{r,\nu},\theta_\nu)$ in Eq.~\eqref{eq:events}, and $R^e_e(E_{r,\nu}, E_\nu)$ and $R^e_\theta(\theta_{r,\nu},\theta_\nu)$ in Eq.~\eqref{eq:eventse}, are taken to be gaussian functions. The assumed gaussian widths $\sigma_E/E$ and $\sigma_\theta$ for $\nu_e$, $\bar{\nu}_e$, $\nu_\mu$ and $\bar{\nu}_\mu$ charged-current interactions on argon are shown in Fig.~\ref{fig:resolutions}. We use the dashed curves in the figure to parametrize the resolutions as functions of the neutrino energy $E_{\nu}$, according to:
\begin{equation}
  \begin{split}
    \sigma_E/E_{\nu} = A/E_{\nu}^B ~,  \\
    \sigma_\theta = C/E_{\nu} + D ~.
  \end{split}
  \label{eq:resolutions}
\end{equation}

The numerical values for the parameters in  Eq.~\ref{eq:resolutions} are reported in Appendix~\ref{sec:appendix}. The resolution functions were obtained via fast Monte-Carlo simulations as follows, similarly to what was done in \cite{Acciarri:2015uup,Adams:2013qkq}. First, large samples of mono-energetic neutrino-argon interactions are simulated with GENIE, for the various neutrino flavors ($\nu_e$, $\bar{\nu}_e$, $\nu_\mu$ and $\bar{\nu}_\mu$) and for the relevant neutrino energy range 0.5--8~GeV. The GENIE simulation includes nuclear effects. Second, for each event, each final state particle exiting the nucleus has its kinetic energy and angular direction smeared according to the assumptions described in \cite{Adams:2013qkq}. The relative energy resolutions are taken to be $1\%/\sqrt{E_e}+1\%$, $3\%$ and $30\%/\sqrt{E_\text{had}}$ for electrons, muons and hadrons, respectively, where $E_e$ and $E_\text{had}$ are expressed in GeV. The absolute angular resolutions are taken to be $1^\circ$, $1^\circ$ and $10^\circ$ for the same three final state particle categories and  for all energies. Third, the incoming neutrino energy and direction of each interaction is reconstructed as follows:
\begin{equation}
  \begin{split}
    E_{r,\nu} = K_{r,l} + m_l + \sum_h K_{r,h} ~, \\
    \theta_{r,\nu} = \arccos(p_{zr,\nu} / \lvert\vec{p}_{r,\nu}\rvert) ~,
  \end{split}
  \label{eq:recoquantities}
\end{equation}
where $K_{r,l}$ and $K_{r,h}$ are the reconstructed charged lepton and hadron kinetic energies, $m_l$ is the charged lepton mass, the sum $\sum_h$ is intended over all final state hadrons, and $\vec{p}_{r,\nu}\equiv \vec{p}_{r,l}+\sum_h \vec{p}_{r,h}$  is the reconstructed 3-momentum of the incoming neutrino, where the true neutrino direction is defined along the z axis. Fourth, histograms of the reconstructed neutrino energy and direction are obtained for each (neutrino flavor, neutrino energy) simulated data sample. Fifth, $\sigma_E/E_{\nu}$ and $\sigma_\theta$ for each data sample are obtained from a gaussian fit to the energy  histogram and from the mean of the angle histogram,  respectively. The resolution functions are shown in Fig.~\ref{eq:resolutions} for each sample via marker symbols. Sixth, the energy dependence of the resolutions functions is parametrized according to Eq.~\eqref{eq:resolutions}.

The behavior of the resolution functions in Fig.~\ref{eq:resolutions} can be easily understood. The main effect is that both $\sigma_E/E$ and $\sigma_\theta$ improve noticeably as the neutrino energy increases. For $\sigma_E/E$, this is a direct consequence of the relative energy resolutions assumed for electrons and (especially) hadrons, improving as the particle energies increase. For $\sigma_\theta$, this is due to the Fermi momentum of the target nucleon, whose angular smearing effect is more important at low neutrino energies. A second, smaller, effect can also be appreciated in Fig.~\ref{eq:resolutions}, namely that antineutrino resolutions are slightly better than neutrino ones, both for $\sigma_E/E$ and $\sigma_\theta$. On the one hand, this is due to the fact that the average inelasticity (or energy fraction carried away by final state hadrons) is somewhat lower in antineutrino interactions \cite{Friedland:2018vry}, and on  the other because hadron resolutions are substantially worse than charged lepton ones. An even smaller difference can be appreciated between the relative energy resolutions of electron and muon antineutrinos of the same energy. In this case, electron antineutrino energy resolutions are slightly better, because of the better assumed accuracy in reconstructing electron energy ($1\%/\sqrt{E_e}+1\%$) compared to muon energy (3\%).

Our energy resolution assumptions in Fig.~\ref{fig:resolutions} are similar to the ones in \cite{Acciarri:2015uup,Adams:2013qkq}, that use similar methodologies and assumptions. They are qualitatively similar also to the ones obtained in more recent studies, see Refs.~\cite{DeRomeri:2016qwo,Friedland:2018vry}. On the other hand, we are not aware of other neutrino angular resolutions studies in LArTPCs to compare our findings with.


\section{Analyses and Results}
\label{sec:analysis}

Here, we describe the statistical analysis and how we extract the
sensitivity to the neutrino mass ordering. In order to emphasize the
impact of the muon capture in
Argon, we present two possible analyses. 
The first case will assume that no charge identification is possible. Then, we will focus on the extra
bonus that the muon capture in Argon process provides. 

In the following, we define a fiducial mass ordering, 
\emph{true ordering (TO)}, in order to generate mock data. 
Then, we try to reconstruct the event rates using the 
\emph{wrong ordering (WO)} assumption. Although there is some 
preference for normal neutrino mass ordering, as previously stated,
we shall study also the case of inverted ordering as TO.

\begin{table}[t!]
\centering
\begin{tabular}{|c|c|c|}
\hline  
parameter & Normal ordering & Inverted Ordering 
\\
\hline
$\Delta m^2_{21}$& $7.55\times 10^{-5}$ eV$^2$& $7.55\times 10^{-5}$ eV$^2$\\  
$\Delta m^2_{31}$&  $2.50\times 10^{-3}$ eV$^2$& $-2.42\times 10^{-3}$ eV$^2$\\
$\sin^2\theta_{12}$ & 0.320 & 0.320\\  
$\sin^2\theta_{13}$ & 0.02160 & 0.0222\\
$\delta$ & 0 & 0 \\
\hline
\end{tabular}
\caption{The oscillation parameters used to generate the mock data~\cite{deSalas:2017kay}. We use various values for the atmospheric angle $\theta_{23}$.}
\label{tab:params} 
\end{table}

We use Eqs.~(\ref{eq:events}) and~(\ref{eq:eventse}) to generate our mock
data, using the oscillation parameters from
Tab.~\ref{tab:params} and assuming a 400~kt$\cdot$yr exposure. We will present our results as a function of the atmospheric angle $\theta_{23}$. Therefore, there is no fixed value for this angle in the table. Notice that, since our main sensitivity comes
from the $\nu_\mu\rightarrow\nu_\mu$ channel, the effects of the CP
violating phase $\delta$ are negligible and therefore we set  $\delta
= 0$, finding very similar results for other values of the CP phase. 

Next, we  try to reconstruct the event rates following the two methods mentioned above. 
Before presenting our results, let us discuss our treatment of
systematic uncertainties. We consider several sources of systematic uncertainties in our analyses, coming from
the fact that we do not have a perfect knowledge of the atmospheric
flux and detector response.
In particular, we include an overall rate normalization error accounting for both flux normalization and detector efficiency uncertainties, an error on the 
$\nu$/$\bar{\nu}$ atmospheric flux ratio, and an error on the $\nu_\mu$/$\nu_e$ atmospheric flux ratio.
We follow Ref.~\cite{Adams:2013qkq} and assume a 15\%, 5\%, and 2\% gaussian error on these three quantities, respectively.
We have verified that adding a systematic on the spectral index of the neutrino flux would have a negligible effect. As explained in the previous section, we also add an uncertainty on $\epsilon^\text{cap}$, see Eq.~\eqref{eq:muminuscapture}.
Apart from the systematic uncertainties, we also marginalize over the
oscillation parameters $\Delta m_{31}^2$, $\sin^2\theta_{13}$ and $\sin^2\theta_{23}$ within their current 3$\sigma$ ranges for both orderings, namely $|\Delta m_{31}^2| \in [2.31, 2.60]\times 10^{-3}$ eV$^2$, $\sin^2\theta_{13} \in [0.0196, 0.0244]$ and $\sin^2\theta_{23} \in [0.455, 0.599]$. It is well known
that the solar parameters do not have big effects in atmospheric
neutrino oscillations, hence, they are fixed to their best-fit values
throughout the analysis. 

\subsection*{Method A: Analysis without muon capture tagging}

In this case, muons and antimuons cannot be distinguished. We therefore build  a $\chi^2$ function in the following way: 
\begin{equation}
\chi^2_A(\sin^2\theta_{23}^\text{true})
=
\min_{\text{sys},\Delta m_{31}^2,\theta_{13},\theta_{23}}\left\{\chi^2_{\mu^- + \mu^+} + \chi^2_{e^- + e^+}\right\}\,.
\end{equation}
We use a Poissonian $\chi^2$, which for muons is
\begin{equation}
\chi^2_{\mu^- + \mu^+} = 2 \sum_{i,j} N^\text{WO}_{i,j,\mu} - N^\text{TO}_{i,j,\mu} + N^\text{TO}_{i,j,\mu}\log\left(\frac{N^\text{TO}_{i,j,\mu}}{N^\text{WO}_{i,j,\mu}}\right), 
\end{equation}
where $N^\text{TO(WO)}_{i,j,\mu} = N^\text{TO(WO)}_{i,j,\mu^+} + N^\text{TO(WO)}_{i,j,\mu^-}$ is the sum of the muon and antimuon contributions. 
The same formula applies to $\chi^2_{e^- + e^+}$,  with the replacement $\mu\rightarrow e$. 
The results of our analysis with method A are shown as red curves in Fig.~\ref{fig:result}. 
Note that the sensitivity ranges between 1.5 and 3.5$\sigma$
approximately when normal ordering is the TO (solid lines) and between 1.5 and 2$\sigma$ for true inverted ordering (dashed line).

\subsection*{Method B: Analysis with muon capture tagging}
\begin{figure}
\centering
\includegraphics[width=0.5\textwidth]{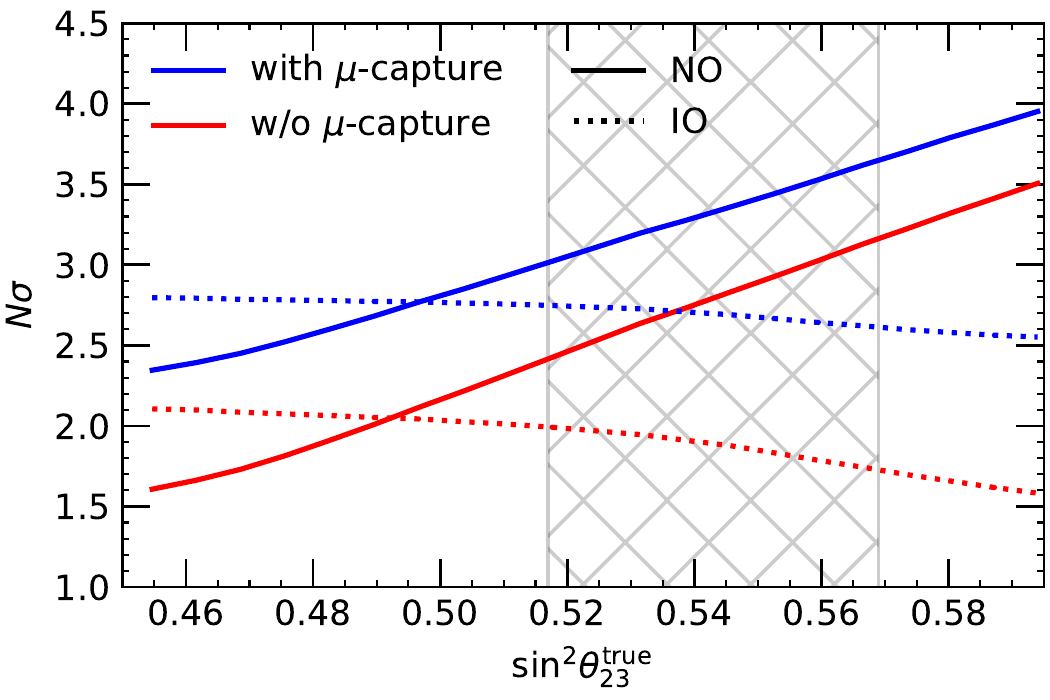}
\caption{The DUNE sensitivity to the neutrino mass ordering as a function of $\sin^2\theta_{23}^\text{true}$. Red (blue) lines correspond to the analysis method A (B). Solid lines are for normal ordering as true ordering, while dashed lines show the sensitivity in the case of true inverted ordering.
The grey band corresponds to the current 1$\sigma$ region for the atmospheric angle.}
\label{fig:result}
\end{figure}

In this other strategy, we use use muon capture to distinguish $\sim$72\% of the muons from antimuons. Therefore, this time our $\chi^2$ function contains three terms, namely
\begin{equation}
\chi^2_B(\sin^2\theta_{23}^\text{true})
= \min_{\text{sys},\Delta m_{31}^2,\theta_{13},\theta_{23}}\left\{\chi^{2,\text{cap}}_{\mu} + \chi^{2,\text{rest}}_{\mu} + \chi^2_{e^- + e^+}\right\}\,.
\end{equation}
The electron term is the same as for method A, while  the other two terms, corresponding to the events with muon capture (cap) and all other events (rest), are given by
\begin{equation}
\chi^{2,X}_{\mu} = 2 \sum_{i,j} N^{\text{WO},X}_{i,j,\mu} - N^{\text{TO},X}_{i,j,\mu} + N^{\text{TO},X}_{i,j,\mu}\log\left(\frac{N^{\text{TO},X}_{i,j,\mu}}{N^{\text{WO},X}_{i,j,\mu}}\right), 
\end{equation}
where $X\in\{\text{cap, rest\}}$, see Eqs.~\eqref{eq:Ncap} and \eqref{eq:Nrest}
. The results of the analysis with
muon capture are shown in Fig.~\ref{fig:result} by the  blue curves. 
As before, true normal ordering is shown as a solid line, while the case of true inverted ordering is represented by a dashed line. 
 The grey band in the figure represents the current 1$\sigma$ allowed region for $\sin^2\theta_{23}$.
Note how the sensitivity to the mass ordering is now at the 2.5--4$\sigma$ level, implying an important improvement with respect to  the results obtained with method A. 
In particular, for the current best-fit point~\cite{deSalas:2017kay}
we find that, using atmospheric neutrinos with muon capture, 
DUNE could measure the neutrino mass ordering at the $3.5 \sigma$ level.
Our method B results can also be compared with the results in the DUNE Conceptual Design Report~\cite{Acciarri:2015uup}, where a similar sensitivity reach and dependence on $\sin^2\theta_{23}$ were obtained.
Compared to~\cite{Acciarri:2015uup}, however, our results more clearly highlight the importance of the muon capture tag.

\section{Conclusions}
\label{sec:concl}

We have explored the advantages of muon capture on Argon nuclei, a process that improves the sensitivity to the neutrino mass ordering using atmospheric  neutrino events at the Liquid Argon Time Projection Chamber DUNE far detector. This is a very relevant result, since it comes without any extra cost. Furthermore,  it can be combined with DUNE beam neutrino results, allowing for an enhancement in the total sensitivity to the mass ordering determination. It is important to notice that our results are applicable to any experiment using Argon. In the case of accelerator-based neutrinos, where significant $\nu_\mu$ contamination exists in the $\bar{\nu}_\mu$ beam, statistical neutrino/antineutrino separation based on muon capture could also be used to enhance DUNE oscillation sensitivities.  

\begin{acknowledgments}
This research made use of the \texttt{MINUIT} algorithm\cite{1975CoPhC..10..343J} via the \texttt{iminuit}\cite{iminuit} Python interface.
We would like to thank Peter Denton for useful comments on the draft. 
Work supported by the Spanish grants
FPA2015-68783-REDT, 
FPA2017-90566-REDC (Red Consolider MultiDark),
FPA2017-85216-P, 
FPA2017-85985-P, 
FPA2017-82081-ERC
and
SEV-2014-0398 
(AEI/FEDER, UE, MINECO),
and PROMETEO/2018/165 (Generalitat Valenciana).
SG receives support from the European Union's Horizon 2020 research and innovation programme under the Marie Sk{\l}odowska-Curie individual grant agreement No.\ 796941.
OM and MS are also supported by the European Union's Horizon 2020 research and innovation program under the Marie
Sk\l odowska-Curie grant agreements No.\ 690575 and 674896.
CAT is supported by the MINECO fellowship BES-2015-073593.
MT acknowledges financial support from MINECO through the Ram\'{o}n y Cajal contract RYC-2013-12438.
\end{acknowledgments}


\appendix
\section{Neutrino energy and angle resolutions}
\label{sec:appendix}

Figure~\ref{fig:resolutions} shows our estimated neutrino energy $\sigma_E/E_{\nu}$ (top) and neutrino angle $\sigma_\theta$ (bottom) resolutions as a function of neutrino energy $E_{\nu}$, for charged-current  neutrino interactions on argon. The resolutions are parametrized according to Eq.~\ref{eq:resolutions}. The A and B parameters describing the relative neutrino energy resolution (in percent), and the C and D parameters describing the neutrino angle resolution (in degrees), are given in Tabs.~\ref{tab:energyres} and \ref{tab:angleres}, respectively. The parameters are given separately for each neutrino flavor: $\nu_e$, $\bar{\nu}_e$, $\nu_{\mu}$ and $\bar{\nu}_{\mu}$. The parameters in Eq.~\ref{eq:resolutions} assume that $E_{\nu}$ is expressed in GeV.

\begin{table}[t!]
\centering
\begin{tabular}{|c|c|c|c|c|}
\hline  
Parameter & $\nu_e$ & $\bar{\nu}_e$ & $\nu_{\mu}$ & $\bar{\nu}_{\mu}$ 
\\
\hline
A & 22.4  & 20.8  & 22.0  & 20.3 \\
B & 0.582 & 0.680 & 0.548 & 0.625  \\
\hline
\end{tabular}
\caption{Numerical values for the parameters appearing in Eq.~\ref{eq:resolutions}, and defining the energy dependence of the neutrino energy resolution assumed in this work.}
\label{tab:energyres} 
\end{table}

\begin{table}[t!]
\centering
\begin{tabular}{|c|c|c|c|c|}
\hline  
Parameter & $\nu_e$ & $\bar{\nu}_e$ & $\nu_{\mu}$ & $\bar{\nu}_{\mu}$ 
\\
\hline
C & 7.85  & 8.42  & 7.79  & 8.46 \\
D & 3.70 & 2.31 & 3.71 & 2.29  \\
\hline
\end{tabular}
\caption{Numerical values for the parameters appearing in Eq.~\ref{eq:resolutions}, and defining the energy dependence of the neutrino angle resolution assumed in this work.}
\label{tab:angleres} 
\end{table}

\bibliographystyle{kp}

\begingroup\raggedright\endgroup

\end{document}